\documentclass[aps,preprintnumbers,eqsecnum,amsmath,amssymb,nofootinbib]{revtex4}  
\usepackage{graphicx} 
\usepackage{bm}

\begin{document}
\title{Theoretical analysis of the leptonic decays $\bm{B\to \ell \ell \ell\bar\nu_{\ell}}$:\\
  Identical leptons in the final state}
\author{Mikhail A.~Ivanov$^a$ and Dmitri Melikhov$^{a,b,c}$}
\affiliation{
$^a$Joint Institute for Nuclear Research, Dubna, 141980, Russia\\
$^b$D.~V.~Skobeltsyn Institute of Nuclear Physics, M.~V.~Lomonosov Moscow State University, 119991, Moscow, Russia\\
$^c$Faculty of Physics, University of Vienna, Boltzmanngasse 5, A-1090 Vienna, Austria}
\date{\today}
\begin{abstract} 
We study the effects of the identical leptons in the final state of the $B^+\to \ell^+ \ell^- \ell^+\bar\nu_{\ell}$ decay.
The amplitude of the process is described by the same form factors as the amplitude of the
${B\to \ell \ell \ell'\bar\nu'_{\ell}}$ decay for non-identical leptons in the final state. However,
the differential distributions are strongly different, as the
${B^+\to \ell^+ \ell^- \ell^+\bar\nu_{\ell}}$ amplitude contains both the direct ($M_a$) and the exchange ($M_b$) diagrams.
We calculate a number of the differential distirbutions. In particular, we propose an interesting observable that can be
readily measured experimentally --
the differential distribution over the invariant mass of the pair of leptons of the same charge, $\ell^+ \ell^+$.
The good news is that the interference between $M_a$ and $M_b$, $d{\cal B}_{ab}$, is found to be
at the level of less than 1\% in all considered differental distributions 
and therefore can be neglected in the full kinematical region of this decay. 
\end{abstract}
\maketitle
\section{Introduction}
\label{Sec_introduction}
This paper extends our recent analysis \cite{mi2022} of the $B\to lll'\nu'$ decay ($l`\ne l$) to the case of the
identical leptons in the final state ($l'=l$). Such reactions are being studied experimentally
\cite{exp1,exp2,exp3,exp4}, thus requiring a proper theoretical understanding.
By now, there have been a few theoretical papers \cite{sehgal,nikitin,bharucha2021,beneke2,wang2022}, where $B$-decays into two
lepton pairs have been studied.

\begin{figure}[b!]
\includegraphics[height=3.7cm]{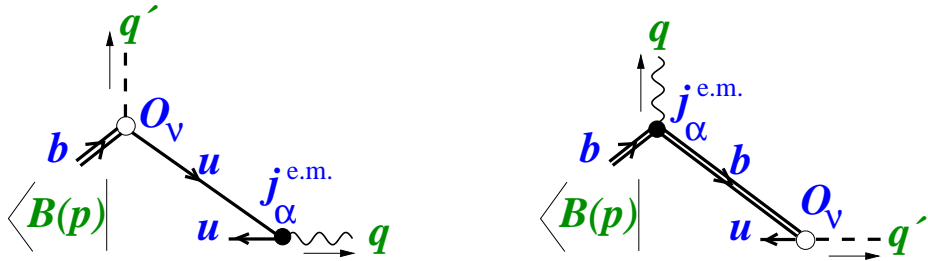}
\caption{\label{Fig:1} 
Feynman diagrams describing the amplitude (\ref{def}).}
\end{figure}

The $B\to \gamma^*l'\nu'$ amplitude (see Fig.~\ref{Fig:1}) may be parametrized via Lorenz-invariant
form factors as follows: 
\begin{eqnarray}
\label{def}
T_{\alpha\nu}(q,q'|p)=i\int dx\,  e^{i q x} \langle 0| T\{j^{\rm e.m.}_\alpha(x),\bar u(0){\cal O_\nu}b(0))\}|\bar B_u(p)\rangle = 
\sum_i L^{(i)}_{\alpha\nu}(q,q')F_i(q'^2,q^2)+\ldots, \quad p=q+q', 
\end{eqnarray}
with $q'$ the momentum of the weak $b\to u$ current, 
and $q$  the momentum of the electromagnetic current.
In Eq.~(\ref{def}), ${\cal O_\nu}=\gamma_\nu, \gamma_\nu\gamma_5$ and $j_\alpha^{\rm e.m.}$
is the conserved electromagnetic current 
\begin{eqnarray}
\label{jem}
j_\alpha^{\rm e.m.}(0)=e Q_b \bar b(0)\gamma_\alpha b(0) + e Q_u \bar u(0) \gamma_\alpha u(0).
\end{eqnarray} 
The quantities $L^{(i)}_{\alpha\nu}(q,q')$ represent the transverse Lorents 
structures, $q^\alpha L^{(i)}_{\alpha\nu}(q,q')=0$, and the dots stand for the longitudinal part which is constrained by the 
conservation of the electromagnetic current, $\partial_\alpha j_\alpha^{\rm e.m.}=0$,
and the equal-time commutation relations. 

The form factors $F_i(q'^2,q^2)$ are complicated functions of two variables, $q'^2$ and $q^2$;
the general properties of these objects in QCD have been studied recently in \cite{ims}.
Noteworthy, gauge invariance provides essential constraints on some of 
the form factors describing the transition of the $B$-meson into the real photon, i.e.,
at $q^2=0$ \cite{bmns2001,m2002,kruger,kmn2016}.

In the past, theoretical analyses focused on a family of similar reactions, namely, 
the $B\to\gamma l^+l^-$ and $B\to \gamma l\nu$ decays
(see, e.g., \cite{aliev,korchemsky,kou,mn2004,beneke1,kmn2018,beneke2018,ivanov,zwicky2019,bobeth,zwicky2021}); 
these processes are described by the same form factors as four-lepton $B$-decays,
but evaluated at a zero value of one of the momenta squared. 
The corresponding form factors depend on one variable $q'^2$, $q'$ the momentum of the weak current; for instance,
for radiative leptonic decays $B\to \gamma l'\nu'$, one needs the form factors $F_i(q'^2,q^2=0)$. 

The four-lepton decay of interest, $B\to l^+l^-l'\nu'$, requires the form factors $F_i(q'^2,q^2)$ for $0<q^2,q'^2<M_B^2$. 
The dependence of the form factors on the variable $q'^2$ can be predicted reasonably well: 
there are no hadron resonances in the full decay region $0<q'^2<M_B^2$, and the $q'^2$-dependence of the form factors 
is determined to a large extent by the influence of the beauty mesons with the appropriate quantum numbers; all these 
mesons are heavier than the $B$-meson and therefore lie beyond the physical decay region of the variable $q'^2$. 
The calculation of the $q^2$-dependence of the form factors is a much more difficult task: light vector mesons
$V=\rho^0,\omega,...$ lie 
in the physical decay region and should be properly taken into account. At $q^2$ in the region
of light vector meson resonances, the form factors cannot be obtained directly in pQCD \cite{ims}.
Here considerations based on the explicit account of these light 
vector resonances---including their finite width effects---are mandatory; the resonance contributions of interest 
may be unambiguously expressed via the weak $B\to V$ form factors.
Then, at $q^2=0$, gauge-invariance constrains the values of the form factors. 
These features allow us to calculate the form factors $F_i(q'^2,q^2)$ in the region $0<q^2\le 1-2$ GeV$^2$,
which dominates the four-meson decay rates and obtain consistent predictions for the latter.

A relatively simple case of different lepton flavours $l\ne l'$ was considered in our recent paper \cite{mi2022}.  
In that case one can easily calculate the differential distribution in $q^2$, $q$ the momentum of the $l^+l^-$ pair,
as well as in $q'^2$, $q'$ the momentum of the $l'\nu'$ pair: the angular variables do not enter the form factors
and as the result all angular integrations may be calculated explicitly yielding the differential distributions
in $q^2$ and in $q'^2$ in explicit forms.

This paper focuses on the case of the identical leptons in the final state $l=l'$. The amplitude is described by the
same form factors as the case $l\ne l'$, so we use the model for these form factors constructed in \cite{mi2022}.
However, a specific feature of the case of the identical leptons is the appearance of the exchange diagrams.
For such diagrams, the variables $q^2$ and $q'^2$ that determine the form factors do not coincide
with the momenta of the $l^+l^-$ and $l^+\nu$ pairs in the final state and thus the angular variables appear explicitly
in the form factors. As the reuslt, the contribution of the exchange diagrams cannot
be obtained as an explicit analytic expression and a numerical evaluation of the phase-space integrals is necessary.
We provide here all necessary details for the theoretical description of this reaction and report the numerical
predictions for a number of the differential distributions.

We propose an interesting kinematical variable: the differential distribution in the momentum of
the pair of the same-charge leptons (i.e. the $l^+l^+$ lepton in the case of the $B^+\to l^+l^-l^+\nu$
decay and the $\mu^+e^+$ pair in the case of the $B^+\to \mu^+\mu^- e^+\nu_e$ or the
$B^+\to e^+e^- \mu^+\nu_\mu$ decay).
This distribution can be measured experimentally in a straightforward way and we obtain predictions for this
differential distribution.


\section{\label{Sect4} The $B^-\to l^+l^-l'^-\bar \nu'$ form factors}
The amplitude of the $B\to ll l'\nu'$ transition for $l'\ne l$ may be parameterized as follows
(see also \cite{Lattice2021}) :
\begin{eqnarray}
A(B\to ll l'\nu')= i e^2 \frac{G_F}{\sqrt2}\,V_{ub}\cdot 
\bar l\gamma_\alpha l \cdot  \bar l'\gamma_\nu (1-\gamma_5)\nu' \frac{1}{q^2}
\bigg\{
(g_{\alpha\nu}q'q-q'_\alpha q_\nu)\frac{F_{1A}}{M_B}
+q'_\alpha q_\nu\frac{F_{2A}}{M_B}+q'_\alpha q'_\nu\frac{F'_{2A}}{M_B}
+ i \epsilon_{\nu\alpha q'q}\frac{F_{V}}{M_B}
\bigg\}, 
\nonumber\\
\end{eqnarray}
where the form factors satisfy the following constraints
\begin{eqnarray}
\label{F2A}
F_{2A}(q'^2,q^2=0)&=&0, \\
\label{Fprime2A}
F'_{2A}(q'^2,q^2=0)&=&\frac{2Q_Bf_BM_B}{M_B^2-q'^2}. 
\end{eqnarray}
Explicit formulas for the differential distributions in the case $l'\ne l$ have been derived in
\cite{mi2022}; we do not repeat these formulas here but refer to \cite{mi2022}.  

The same form factors parameterize the amplitude for the case $l'\ne l$; however, one has to take into account
the contribution of the lepton exchange diagrams in which the variables $q^2$ and $q'^2$ have a complicated
relationship with the momenta of the final lepton pairs. The details are given in the next Section.
We now remind the essential features of our model of the form factors as developed in \cite{mi2022}. 

\noindent
$\bullet$
The contribution of the form factor $F'_{2A}(q^2,q'^2)$ can be neglected in the case $l=l'$
so in what follows we neglect its contribution.

\noindent
$\bullet$ For the form factors $F_{1A,2A,V}(q'^2,q^2)$ we use single-subtracted dispersion representations in $q^2$.
This allows us to take into account all constraints coming from gauge invariance and from the known behaviour in the
large-energy limit of QCD \cite{korchemsky}.

\noindent
$\bullet$ We assume that the spectral densities are saturated by light vector-meson
resonances $\rho^0$ and $\omega$ in the $q^2$-channel. Since these resonances emerge in the physical region of
the $B$-decay of interest, we take into account the $q^2$-dependent finite widths of these
resonances \cite{nachtmann}. In the end, we come to the following expressions for the form factors
\begin{eqnarray}
\label{FA1}
F_{1A}(q'^2,q^2)&=&F_{A}(q'^2)-\frac{Q_B f_B M_B}{q'q}-
q^2
\sum\limits_{V=\rho^0,\omega}\bigg(
\frac{1}{M_V^2}\frac{2M_B(M_B+M_V)}{M_B^2-M_V^2-q'^2}\frac{M_V f_V}{M_V^2-q^2-i\Gamma_V(q^2)M_V}A_1^{B\to V}(q'^2)
\bigg),\nonumber\\
\\
\label{FA2}
F_{2A}(q'^2,q^2)&=&-q^2 M_B
\sum\limits_{V=\rho^0,\omega}
\frac{1}{M_V^2}\frac{2 M_V f_V}{M_V^2-q^2-i\Gamma_V(q^2)M_V}
\left[
\frac{M_B+M_V}{M_B^2-M_V^2-q'^2}A_1^{B\to V}(q'^2)-\frac{A_2^{B\to V}(q'^2)}{(M_B+M_V)}
\right]
\nonumber\\
&&+Q_Bf_B\left(\frac{2M_B}{M_B^2-q'^2}-\frac{2M_B}{M_B^2-q'^2-q^2}\right),
\\
\label{FV}
F_V(q'^2,q^2)&=&F_V(q'^2)-q^2M_B
\sum\limits_{V=\rho^0,\omega}\bigg(
\frac{1}{M_V^2}
\frac{M_V f_V}{M_V^2-q^2-i\Gamma_V(q^2)M_V}\frac{2 V^{B\to V}(q'^2)}{M_B+M_V}\bigg). 
\end{eqnarray}
\noindent 
$\bullet$ 
The form factors $F_A(q'^2)$ and $F_V(q'^2)$ describe the $B\to \gamma l'\nu'$ transition; 
they emerge as subtraction terms at $q^2=0$ in the $q^2$-disperison representations for the form factors 
$F_{1A,V}(q'^2,q^2)$. The form factors $F_A(q'^2)$ and $F_V(q'^2)$ 
are equal to each other at the leading order of the double $1/E_\gamma$ ($2 M_B E_\gamma=M_B^2-q'^2$) and 
$1/M_B$ expansions in QCD \cite{korchemsky} but differ in the subleading orders \cite{mn2004,beneke1,beneke2018}:  
\begin{eqnarray}
\label{leet1}
F_{A}(q'^2)&=&-\frac{Q_u f_B M_B}{2 E_\gamma\lambda_B}+\frac{Q_b f_B M_B}{2 E_\gamma m_b}+O(Q_u f_B M_B/E_\gamma^2), \\
\label{leet2}
F_{V}(q'^2)&=&-\frac{Q_u f_B M_B}{2 E_\gamma\lambda_B}-\frac{Q_b f_B M_B}{2 E_\gamma m_b}+O(Q_u f_B M_B/E_\gamma^2)
\end{eqnarray}
The magnitude of the form factors $F_A(q'^2)$ and $F_V(q'^2)$ is determined to a large extent by the
parameter $\lambda_B$, the inverse moment of the $B$-meson light-cone distribution amplitude $\phi_B$
\cite{korchemsky}. Taking into account a large
uncertainty in the present knowledge of the parameter $\lambda_B$
\cite{beneke1,kou,BraunIvanovKorchemsky2004,kmn2018,zwicky2021,mn2004}, we use the
monopole forms (\ref{leet1}) and (\ref{leet2}) in the full
kinematically allowed region of $q'^2$ and allow the variation of $\lambda_B$ in the range
$\lambda_B(1\,{\rm GeV})=(0.5\pm 0.15)$ GeV. 

\noindent  
$\bullet$
The contributions of the light vector mesons $V=\rho^0,\omega$ to the form factors $F_{1A,2A,V}(q'^2,q^2)$
are {\it unambiguous} (cf. \cite{beneke2}) and are expressed via the form factors
$A_1^{B\to V}(q'^2)$, $A_2^{B\to V}(q'^2)$, and $V^{B\to V}(q'^2)$ 
describing the weak decay $B\to V$. In spite of many efforts to calculate these form factors in a
broad kinematical decay region $0 < q'^2 < M_B^2$, our knowledge of these quantities is not very accurate, 
see e.g. \cite{ms2000,ballzwicky2005,ivanov2,gubernari2019}. For our calculations we use the results from
\cite{ms2000} and assign to them a 10\% uncertainty. 
The uncertainties in these form factors, along with the uncertainty in the parameter $\lambda_B$,
is the second main source of the uncertainty in the theoretical predictions for $B\to l^+l^-l'\nu'$ decays.

The results presented in the next Section are obtained for our form factor model described in full detail
in Section 5 of \cite{mi2022} and for the parameter $\lambda_B=0.65$.

\begin{figure}[h!]
\begin{center}
\vspace*{.25cm} 
\includegraphics[width=12cm]{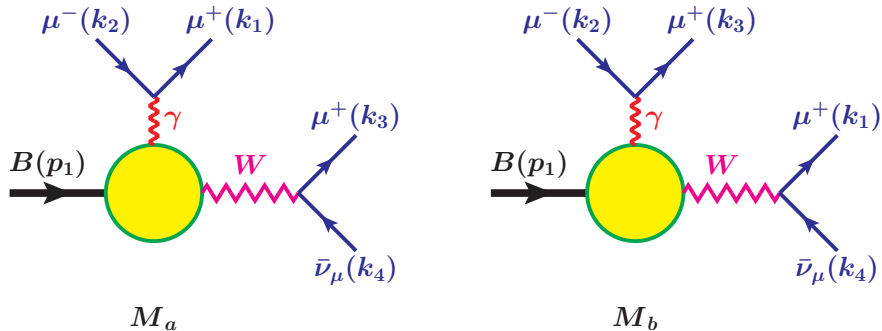}
\caption{Two diagrams describing the $B$-decay into $\mu^+\mu^-\mu^+\bar\nu_\mu$.}
\label{fig:diag}
\end{center}
\end{figure}

\section{The decay \boldmath{$B^+ \to \mu^+\mu^-\mu^+\bar\nu_\mu$}}
The case of two identical leptons is technically more involved than
the case of all different leptons, considered in~\cite{mi2022}. The reason
is that an additional contribution from the interchange of the two $\mu^+$ leptons arises.

The first diagram in Fig.~\ref{fig:diag}, $M_a(k_1,k_2,k_3,k_4)$, 
is the same as for the $B$-decay into non-identical leptons (e.g., $B^+\to \mu^+\mu^- e^+\nu_e$).
The second diagram is obtained from the first one by permutation of two
final identical leptons: $M_b(k_1,k_2,k_3,k_4)= M_a(k_3,k_2,k_1,k_4)$.
The total amplitude for the case of two identical leptons in the final state reads 
\begin{eqnarray}
M_{\rm tot} = \frac{1}{\sqrt{2}}\, (M_a - M_b) \quad\text{and}\quad
|M_{\rm tot}|^2 =
\frac12 \left( |M_a|^2 +  |M_b|^2 - 2\,{\rm Re}(M_a M_b^\ast) \right).
\end{eqnarray}
The factor $1/\sqrt{2}$ in the amplitude corresponds to the factor $1/2$ in the phase space
for the case of two identical leptons. Respectively, we use the expression for the
phase space without the factor $1/2$ correspondng to the identical particles in the final state.


\subsection{The differential distribution over the momentum of the $\mu^+\mu^-$ pair} 
In a theoretical consideration, one can calculate the branching fraction and one-dimensional differential
distribution for two kinematical variables $q_{12}^2=(k_1+k_2)^2$ (momentum of one of the $\mu^+\mu^-$ pairs) 
and $q_{34}^2=(k_3+k_4)^2$ (momentum of the $\mu^+\nu$ pair). For diagram $M_a$, $q_{12}=q$ and $q_{34}=q'$,
so that the angular variables do not enter the form factors; the angular integrals may be taken analytically. 
For diagram $M_b$, the photon momentum $q$ and the
weak-vertex momentum $q'$ do not coincide with $q_{12}$ and $q_{34}$, so that the angular variables appear
in the arguments of the form factors; all angular integrals should be taken numerically. 
\begin{figure}[h!]
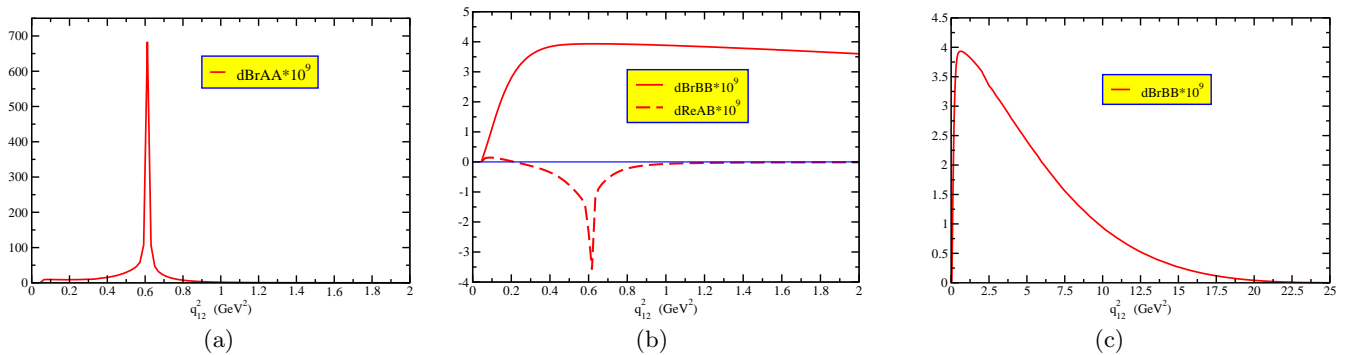

  \centering
  \vspace{.3cm}
  \begin{tabular}{ccc}
    \includegraphics[width=5.4cm]{Plot_q12_dBrAA.eps} \quad &\quad
    \includegraphics[width=5.4cm]{Plot_q12_dBrBB_dReAB_0-2.eps}  \qquad &\qquad
    \includegraphics[width=5.4cm]{Plot_q12_dBrBB.eps}
    \\
    \vspace{.8cm}
    (a) & (b)  & (c)
  \end{tabular}
  \vspace{-.9cm}
  \caption{\label{fig:dif}
    The differential distributions in units $10^{-9}$:
    (a) $d{\cal B}_{aa}(q_{12}^2)$ at $0<q_{12}^2({\rm GeV}^2)<2$; 
    (b) $d{\cal B}_{ab}(q_{12}^2)$ vs $d{\cal B}_{bb}(q_{12}^2)$ at $0<q_{12}^2({\rm GeV}^2)<2$;
    (c) $d{\cal B}_{bb}(q_{12}^2)$ in the full range $4m_\mu^2<  q_{12}^2<(M_B-m_\mu)^2$.}
\end{figure}
Obviously, the contribution to branching fraction coming from $|M_a|^2$ and  $|M_b|^2$ are
identically the same due to
the symmetry $k_1\leftrightarrow k_3$ of the phase-space measure.
But verifying this property is a non-trivial check for numerical evaluation of the five-dimensional
integrals. Fig.~\ref{fig:dif} shows the $q_{12}^2$-differential distributions. The differential distributions over
the variable $q_{23}^2$ ($q_{23}$ the momentum of another $\mu^+\mu^-$ pair that may be isolated in the amplitude)
is the same because of the symmetry of the amplitude: the replacement $k_1\to k_3$ leads to the replacement
$M_a\to M_b$ and vice versa.

\newpage

\subsection{The differential distribution over the momentum of the $\mu^+\nu_\mu$ pair} 
\begin{figure}[h!]
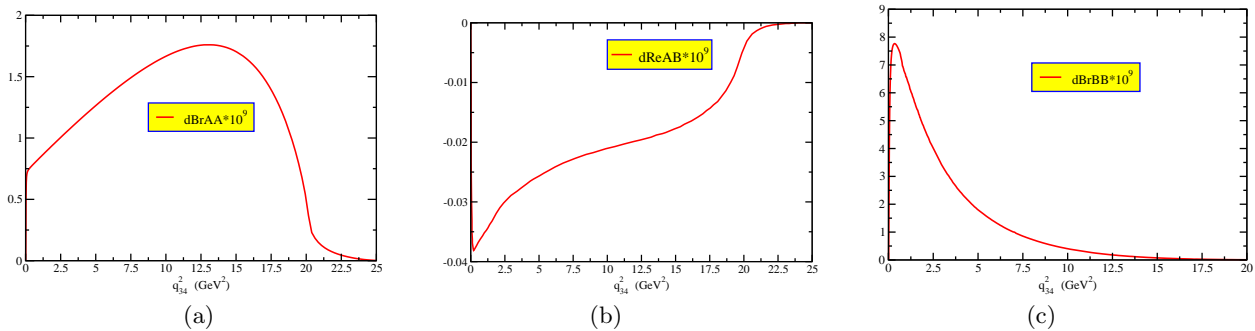

  \centering
  \vspace{.3cm}
  \begin{tabular}{ccc}
    \includegraphics[width=5cm]{Plot_q34_dBrAA.eps}\qquad &\qquad
    \includegraphics[width=5cm]{Plot_q34_dReAB.eps}\qquad &\qquad
    \includegraphics[width=5cm]{Plot_q34_dBrBB.eps}
    \\
    \vspace{.8cm}
    (a) & (b)  & (c)
    \\
  \end{tabular}
  \vspace{-.9cm}
  \caption{\label{fig:diff3}
    The differential distributions in the full range
    $m^2_\mu< q_{34}^2< (M_B-2m_\mu)^2$ (in units $10^{-9}$):
    (a) $d{\cal B}_{aa}(q_{34}^2)$; 
    (b) $d{\cal B}_{ab}(q_{34}^2)$;
    (c) $d{\cal B}_{bb}(q_{34}^2)$}
\end{figure}

In a theoretical consideration, we can also calculate the differential
distribution over $q_{34}^2=(k_3+k_4)^2$ (momentum of the $\mu^+\nu$ pair).
These distributions are shown in Fig. \ref{fig:diff3}. Obviously, the mixed term may be neglected
in the full range of $q_{34}^2$. 

\subsection{The differential distribution over the momentum of the $\mu^+\mu^+$ pair}
\begin{figure}[h!]
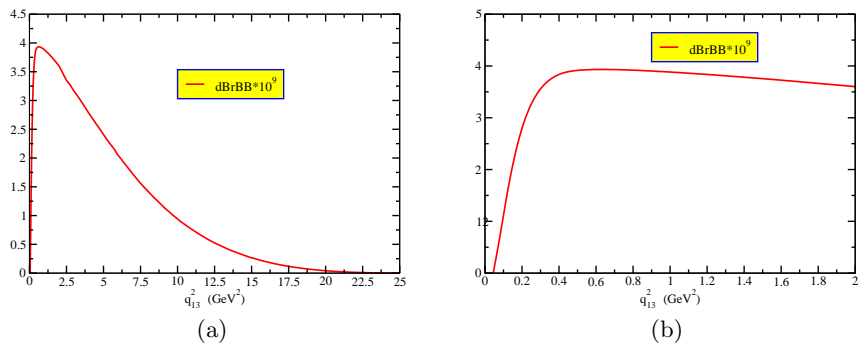

  \centering
  \vspace{.5cm}
\begin{tabular}{cc}
\includegraphics[height=4cm]{Plot_q13_dBrBB.eps}
  \qquad  &  \qquad
\includegraphics[height=4cm]{Plot_q13_dBrBB_0-2.eps}\\
(a)\qquad   & \qquad (b)\\
\end{tabular}
\caption{\label{fig:diff2}
  The differential distribution (in units $10^{-9}$) $d{\cal B}(q_{13}^2)$ over the momentum of the
  same-charge lepton pair $\mu^+\mu^+$ ($q_{13}=k_1+k_3$):
  (a) the full range $4m_\mu^2 <q_{13}^2 < (M_B-m_\mu)^2$; 
  (b) the range $0< q_{13}^2<2$~GeV$^2$.}
\end{figure}

An interesting observable that can be readily measured experimentally is the differential distribution
over the momentum of the $\mu^+\mu^+$ pair. Unlike the $\mu^+\mu^-$ distribitions, one has only one pair
of same-charge leptons in each event. The process is described by the same two diagrams in Fig.~\ref{fig:diag}
but one has to calculate the distribution over the variable $q_{13}^2=(k_1+k_3)^2$. The contributions
$d{\cal B}_{aa}(q_{13}^2)$ and $d{\cal B}_{bb}(q_{13}^2)$ 
are equal to each other and coincide with the distribution $d{\cal B}_{bb}(q_{12}^2)$ discussed above.
Obviously, the mixed $d{\cal B}_{ab}(q_{13}^2)$
term can be safely neglected similar to the case of the distribution in the $l^+l^-$ momentum $q_{12}^2$ considered above: 
(i) the integral $\int d{\cal B}_{bb}(q_{13}^2)dq_{13}^2$ comprises only 1\% of 
$\int d{\cal B}_{aa}(q_{13}^2)dq_{13}^2=\int d{\cal B}_{bb}(q_{13}^2)dq_{13}^2$.
(ii) the distribution $d{\cal B}_{ab}(q_{13}^2)$ contains no resonances and is therefore
smeared over the full kinematical $q_{13}^2$ range as a small addition to
$d{\cal B}_{aa}(q_{13}^2)=d{\cal B}_{bb}(q_{13}^2)$ at the level of 1\%.
Figure \ref{fig:diff2} shows our predictions for $d{\cal B}(q_{13}^2)$. 
\vspace{-.8cm}

\subsection{The branching ratio of the $B^+ \to \mu^+\mu^-\mu^+\bar\nu_\mu$ decay}
Table~\ref{tab:BR} presents our numerical results for the total branching ratio
of the $B^+ \to \mu^+\mu^-\mu^+\bar\nu_\mu$ decay and the separate contributions coming from $|M_a|^2$, $|M_b|^2$ and
the interference term $2\times{\rm Re}(M_aM_b^\ast)$.
We use short-hand notations 
\begin{eqnarray}
{\cal B}_{aa} &=& \frac{\tau_B}{2M_B}\int\!\!d\Phi\, |M_a|^2, \quad
{\cal B}_{bb}  = \frac{\tau_B}{2M_B}\int\!\!d\Phi\, |M_b|^2 , \quad    
{\cal B}_{ab}  = \frac{\tau_B}{2M_B}\int\!\!d\Phi\, 2{\rm Re}(M_aM_b^\ast),
\\
&&
\hspace*{4cm}
{\cal B}_{\rm tot}=\frac12\Big( {\cal B}_{aa} +  {\cal B}_{bb} - {\cal B}_{ab}\Big)
\label{eq:BR}
\end{eqnarray}
where the phase-space measure is given by Eqs.~(\ref{eq:width}) and
(\ref{eq:dPhi}). One can see that the contribution of the interference term
$2\times{\rm Re}(M_aM_b^\ast)$ is negative and two orders of magnitude
less than the contribution of $|M_a|^2$ and $|M_b|^2$.
Thus, the interference term may be neglected. This is a very good news as the calculation
of the interference term is the most time-consuming part of the full calculation. 
\begin{table}[h!]
\centering
\begin{tabular}{|c|c|c|c|c|}
\hline
\quad Mode & $\frac{1}{2}({\cal B}_{aa}+{\cal B}_{bb})$ & $\frac{1}{2}{\cal B}_{ab}$ &
$ {\cal B}_{\rm tot}$ & $ {\cal B}^{\rm exp. cut}_{\rm tot}$
\\
\hline
$\mu^+\mu^-\mu^+\bar\nu_\mu$ &  $2.80\times 10^{-8}$ 
 &  $-2.26\times 10^{-10}$   &  $2.82\times 10^{-8}$  &   $2.73\times 10^{-8}$     
\\
\hline
\end{tabular}
\caption{\label{tab:BR}
Branching ratio of the $B^+ \to \mu^+\mu^-\mu^+\bar\nu_\mu$ decay. 
Separate contributions coming from $|M_a|^2$, $|M_b|^2$ and
the interference term $2\times{\rm Re}(M_aM_b^\ast)$ are also given.
The ${\cal B}^{\rm exp. cut}_{\rm tot}$ is the result
obtained by applying the LHCb event selection criterion (\ref{lhcbcut}).
The results correspond to $\lambda_B=0.65$.}
\end{table}
We also provide the ${\cal B}^{\rm exp. cut}_{\rm tot}$ which is calculated making use of the LHCb \cite{exp4}
event selection criterion: In each event, one can form two $\mu^+\mu^-$ pairs; the events are selected on the basis of the
criterion that the lowest of the two $\mu^+\mu^-$ mass combinations should be less than 0.98 GeV.
In our calculation this corresponds to restricting the phase-space integration by the condition
\begin{eqnarray}
  \label{lhcbcut}
{\rm min}\{(k_1+k_2)^2,(k_3+k_2)^2\}\le 0.96\, {\rm GeV}^2.
\end{eqnarray}

\section{Discussion and Conclusions}
Making use of the model for the form factors of \cite{mi2022}, we performed a detailed
analysis of the exchange diagrams and the interference effects that appear in the case
of the identical leptons in the final state.
We calculate the differential distributions in various variables: namely, in $q_{12}^2$, the square of the invariant
mass of one of the $\mu^+\mu^-$-pairs (Fig.~\ref{fig:dif}), in $q_{34}^2$, the square of the invariant
mass of one of the $\mu^+\nu_\mu$-pairs (Fig.~\ref{fig:diff2}), and in $q_{13}^2$, the square of the invariant
mass of one of the $\mu^+\mu^+$ pair (Fig.~\ref{fig:diff3}).
The latter differential distirbution may be readily measured experimentally. 

Our findings may be summarized as follows: 

$\bullet$
For the differential distribution in $q_{12}^2$, 
$d{\cal B}_{aa}$ has a sharp resonance structure in the region of $\rho$ and $\omega$ resonances.
The distribution of $d{\cal B}_{bb}$ spreads over the full range of $q_{12}^2$ and exhibits no
resonance structure. Nevertheless, the integrated differential rates ${\cal B}_{aa}$ and ${\cal B}_{bb}$
are equal to each other. The interference term $d{\cal B}_{ab}$ contributes at less than 1\% level
and may be safely neglected. Noteworthy, the distribution $d{\cal B}(q_{12}^2)$ is fully determined
by the resonances in all regions of $q_{12}^2$:
In the region $0<q_{12}^2<1$ GeV$^2$ via $d{\cal B}_{aa}$, and in the region of $1\,{\rm GeV}^2 <q_{12}^2$ via
$d{\cal B}_{bb}$. 
Consequently, the perturbative tail of the form factors at $q_{12}^2>1$-$2$ GeV$^2$ does not show up
in the differential distributions for the identical leptons in the final state at all.
This makes an essential difference with the case of non-identical leptons, where the region
$q^2_{12}\gg$ 1 GeV$^2$ is determined by the pQCD behaviour of the form factors. 

$\bullet$
The differential distribution over the momentum of the $\mu^+\nu_\mu$ pair, 
$q_{34}^2$, has an interesting shape, different for $d{\cal B}_{aa}(q_{34}^2)$ and $d{\cal B}_{bb}(q_{34}^2)$,
and a numerically negligible interference term $d{\cal B}_{ab}(q_{34}^2)$. This differential distrubution
is rather interesting theoretically but is unlikely to be measurable experimentally. 

$\bullet$
The differential distribution in $q_{13}^2$, the square of the invariant mass of the $\mu^+\mu^+$ pair, has a relatively
flat non-resonant structure in the full range of $q_{13}^2$. The contribution of the $M_a$ and $M_b$ diagrams are equal to each other,
$d{\cal B}_{aa}(q_{13}^2)=d{\cal B}_{bb}(q_{13}^2)$. 
The interference term $d{\cal B}_{ab}(q_{13}^2)$ is smeared over the full $q_{13}^2$-region
as a minor positive addition at the level of less than 1\% and may be safely neglected. 

$\bullet$ The good news is that the interference term between the direct diagram $M_a$ and the exchanged diagram $M_b$
provides a positive contribution at the level less than 1\% to the differential distribution in all regions
of the kinematical variables and thus can be safely neglected. This greatly simplifies the calculation procedure as
the interference $AB$ term represents the most time-consuming part of the calculations.

$\bullet$ 
For ${\cal B}(B\to \mu^+\mu^-\mu^+\nu_\mu)$, taking into account all uncertianties,
we confirm our result of \cite{mi2022}: 
\begin{eqnarray}
\label{4mu}
      {\rm Br}(B^+\to\mu^+\mu^- \mu^+\bar\nu_\mu)=
       (3.02\;{^{+0.45}_{-0.25}}|_{\lambda_b}\pm 0.62|_{\rm\, weak\, ffs})\, 10^{-8}. 
\end{eqnarray}
Applying the kinematical selection rule for the $\mu^+\mu^-$ pairs (\ref{lhcbcut}) as done by the LHCb collaboration \cite{exp4},
leads to a small reduction at the level of 3\% of our theoretical result (\ref{4mu}).

In summary, we reinforce our previous finding that our theoretical estimate 
is only marginally compatible with the upper limits obtained by the LHCb Collaboration
\cite{exp4} ${\rm Br}(B^+\to\mu^+\mu^- \mu^+\bar\nu_\mu)\le 1.6 \cdot \,10^{-8}$. see a small reduction 

\acknowledgments
We are grateful to M.~Beneke, G.~Gagliardi, L.~Gladilin, S.~Simula, and R.~Zwicky
for valuable and interesting discussions. 
\appendix
\section{Kinematics of the $B$-decay with four leptons in the final state}
\begin{figure}[h!]
\includegraphics[width=8cm]{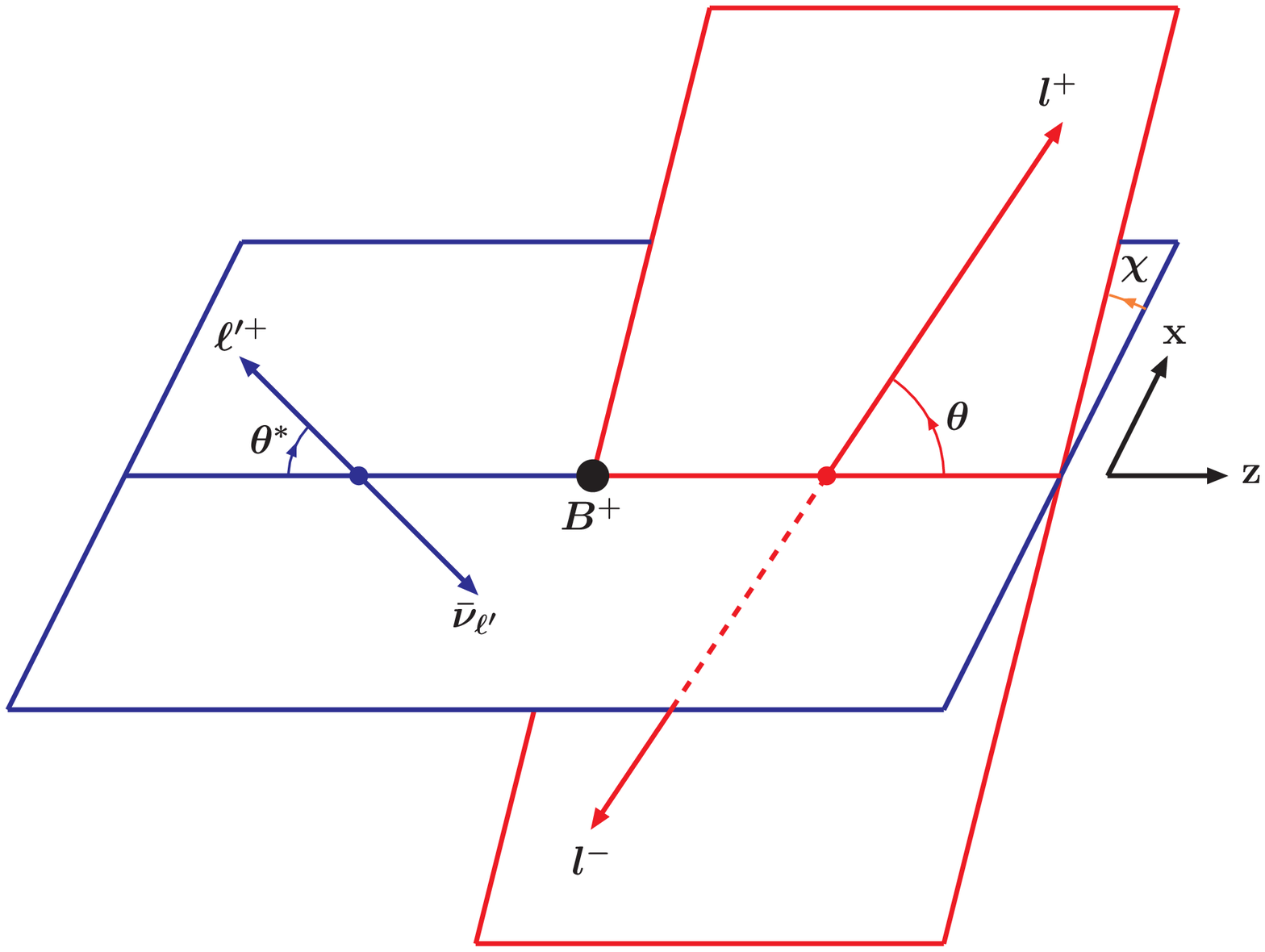}
\caption{\label{Fig:bkangl} 
Definition of angles $\theta^\ast$, $\theta$ and $\chi$ in the decay
of $ B^+ \to \ell^{\prime\,+} + \bar\nu_{\ell'} + \ell^+ + \ell^- $.}
\end{figure}

We consider the reaction
\begin{eqnarray}
B^+(p) \to \ell^{\prime\,+}(k_3) + \bar\nu_{\ell'}(k_4)
+ \ell^+(k_1) + \ell^-(k_2)\,.
\end{eqnarray}
The two planes of the final particles are shown in Fig.~\ref{Fig:bkangl}.
The decay amplitude is described by 5 kinematical variables:
$\bullet$ $q_{34}^2 \equiv (k_3+k_4)^2$ is the $\ell^{\prime\,+}\nu_{\ell'}$ invariant mass;\\
$\bullet$ $q_{12}^2  \equiv (k_1+k_2)^2$ is the dilepton invariant mass;\\
$\bullet$ $\theta^\ast$ is the angle of the $\ell^{\prime\,+}$ in the
  $\ell^{\prime\,+}\bar\nu_{\ell'}$ c.m.s  w.r.t the $\ell^{\prime\,+}\bar\nu_{\ell'}$
  flight direction;\\
$\bullet$  $\theta$ is the angle of the $\ell^+$ in the dilepton c.m.s 
w.r.t  the $\ell^+\ell^-$  flight direction;\\
$\bullet$ $\chi$ is the azimuthal angle between $\ell^{\prime\,+}\nu_{\ell'}$ and dilepton planes.

All particles are on their mass shell: 
\begin{eqnarray}
p^2=M_B^2, \quad 
k_3^2=m_3^2\equiv m_{\ell'}^2, \quad
k_4^2=m_4^2\equiv m_{\nu_{\ell'}}^2 =0 ,\quad 
k_1^2=k_2^2\equiv m_\ell^2.
\end{eqnarray}
We also introduce the mass notations for the momenta squared: $m_{12}\equiv \sqrt{ q_{12}^2}$ and $m_{34}\equiv \sqrt{ q_{34}^2}$.

The boosted 4-momenta from the $\ell^+\ell^-$ c.m.s. to the $B$-meson rest frame are written as
\begin{eqnarray}
q_{12}^\mu &=& ( E_{12},0,0,|{\bf k}|),
\label{eq:frame12->1}\\
k^\mu_1 &=& \tfrac12
                   (E_{12} + v |{\bf k}| \cos\theta,
                        + v m_{12} \sin\theta\cos\chi,
                        + v m_{12} \sin\theta\sin\chi,
             |{\bf k}| + v E_{12} \cos\theta),\\
k^\mu_2 &=& \tfrac12
                   (E_{12} - v |{\bf k}| \cos\theta,
                        - v m_{12} \sin\theta\cos\chi,
                        - v m_{12} \sin\theta\sin\chi,
            |{\bf k}| - v E_{12} \cos\theta).
\end{eqnarray}
Here  $v=\lambda(q_{12}^2,m_l^2,m_l^2)/q_{12}^4 = \sqrt{1-4 m_l^2/q_{12}^2}$ [$\lambda(a,b,c)\equiv (a-b-c)^2-4 bc$] and
\begin{eqnarray}
  |{\bf k}| = \frac{\lambda^{1/2}(M_B^2,q_{34}^2,q_{12}^2)}{2M_B}, \quad
E_{12} = \frac{M_B^2-q_{23}^2+q_{12}^2}{2 M_B}, \quad   
E_{23} = \frac{M_B^2+q_{23}^2-q_{12}^2}{2 M_B}, \quad
E_{12} + E_{23} = M_B. 
\end{eqnarray}
The boosted momenta from the $\ell'\nu_{\ell'}$ c.m.s. to the $B$-meson rest frame read 
\begin{eqnarray}
q_{34}^\mu &=& (E_{34},0,0,-|{\bf k}|),
\label{eq:frame34->1}\\
k^\mu_3 &=& \frac{1}{E_{34}}
                   (E_{34} E_3 + |{\bf k}||{\bf k_3}|\cos\theta^\ast,
                     + E_{34} |{\bf k_3}|\sin\theta^\ast,
                       0,
                     - E_3|{\bf k}| - E_{34}|{\bf k_3}|\cos\theta^\ast ),\\
k^\mu_4 &=& \frac{1}{E_{34}}
                  (E_{34} E_4 - |{\bf k}||{\bf k_3}|\cos\theta^\ast,
                     - E_{34} |{\bf k_3}|\sin\theta^\ast,
                       0,
                     - E_4|{\bf k}| + E_{34}|{\bf k_3}|\cos\theta^\ast \,).
\end{eqnarray}
where
\begin{eqnarray}
|{\bf k_3}| = \frac{\lambda^{1/2}(q_{34}^2,m_3^2,m_4^2)}{2m_{34}}, \quad
E_3 = \frac{q_{34}^2+m_3^2-m_4^2}{2m_{34}}, \quad
E_4 = \frac{q_{34}^2-m_3^2+m_4^2}{2m_{34}}, \quad E_3 + E_4 = m_{34}.
\end{eqnarray}
The differential decay rate is given by
\begin{eqnarray}
&&
d\Gamma(B\to \ell'\nu_{\ell'}\ell^+\ell^-) =
     \frac{1}{2m_1} |M(k_1,\ldots,k_4)|^2\, d\Phi ,\\
&&     
d\Phi =  \frac{1}{(2\pi)^8}\delta^{(4)}(p_1-k_3-k_4-k_1-k_2)\,
           \frac{d^3\vec k_3}{2 k^0_3}\,  \frac{d^3\vec k_4}{2 k^0_4}\,  
           \frac{d^3\vec k_1}{2k^0_1}\,   \frac{d^3\vec k_2}{2k^0_2},
           \qquad 
\label{eq:width}
\end{eqnarray}
where $k^0_i=\sqrt{m_\ell^2 + \vec k_i^{\,2}}$ for $(i=1,2)$ and
      $k^0_i=\sqrt{m_i^2 + \vec k_i^{\,2}}$  for $(i=3,4)$. 

The integration over the phase space may be reduced to the integration
over the two kinematical variables $k^2$ and $q^2$ and three angles
$\theta^\ast$, $\theta$ and $\chi$.
Then, the differential phase volume in Eq.~(\ref{eq:width}) is given by
\begin{eqnarray}
d\Phi = \frac{v}{(4\pi)^6}\frac{|{\bf k}|}{M_B}\frac{|{\bf k_3}|}{E_{34}}\,
dq_{12}2\, dq_{34}^2 \,d\cos\theta^\ast\, d\cos\theta\, d\chi \,
\qquad
0 \le \theta^\ast,\theta\le \pi, \qquad 0\le \chi \le 2\pi.
\label{eq:dPhi}
\end{eqnarray}
The kinematical constrains on the variables $q_{12}^2$ and $q_{34}^2$ come from
the positivity of the $\lambda$-functions:
$\lambda(q_{34}^2,m_{\ell'}^2,0)$, $\lambda(q_{12}^2,m_{\ell}^2,m_{\ell}^2)$ and
$\lambda(M_B^2,q_{12}^2,q_{34}^2)$ and read as
\begin{eqnarray}
4m^2_{\ell}\le q_{12}^2, \quad m^2_{\ell'}\le q_{34}^2, \quad
\sqrt{q_{12}^2}+\sqrt{q_{34}^2}\le M_B^2.
\label{eq:limits}
\end{eqnarray}
For calculating single differential distribution in $q_{12}^2$ or $q_{34}^2$,
we have the following integration limits 
\begin{eqnarray}
dq_{34}^2dq_{12}^2: &&  m_{\ell'}^2\le q_{34}^2 \le (M_B - 2 m_\ell)^2, \quad
4m_\ell^2\le q_{12}^2 \le (M_B-\sqrt{q_{34}^2})^2,\\
dq_{12}^2dq_{34}^2: && 4m_\ell^2\le q_{12}^2 \le (M_B-m_{\ell'})^2, \quad
m_{\ell'}^2\le q_{34}^2 \le (M_B-\sqrt{q_{12}^2})^2,
\label{eq:bounds}
\end{eqnarray}


\begin{thebibliography}{100}
\bibitem{mi2022}
M.~A.~Ivanov and D.~Melikhov,
{\it Theoretical analysis of the leptonic decays $B\to \ell \ell \ell'\bar\nu_{\ell'}$}, 
Phys.~Rev.~D~{\bf 105}, 014028 (2022).
\bibitem{exp1}
LHCb Collaboration (R.~Aaij et al.), 
{\it Search for the rare decay $B_s^0\to \mu^+\mu^-\mu^+\mu^-$}, 
Phys.~Rev.~Lett.~ {\bf 110}, 211801 (2013).
\bibitem{exp2}
ATLAS Collaboration (M.~Aaboud et al.),
{\it Study of the rare decays of $B^0_s$ and $B_0$ into muon pairs
from data collected during the LHC Run 1 with the ATLAS detector}, 
Eur.~Phys.~J.~{\bf C76}, 513 (2016).
\bibitem{exp3}
LHCb Collaboration (R.~Aaij et al.), 
{\it Search for decays of neutral beauty mesons into four muons}, 
JHEP {\bf 1703}, 001 (2017).
\bibitem{exp4}
LHCb Collaboration (R.~Aaij et al.), 
{\it Search for the rare decay $B^+\to \mu^+\mu^-\mu^+\nu_\mu$}, 
Eur.~Phys.~J.~{\bf C79}, 675 (2019).
\bibitem{sehgal}
Y.~Dincer and L.~Sehgal, 
{\it Electroweak effects in the double Dalitz decay $B(s)\to l^+ l^- l'^+ l'^-$}, 
Phys.~Lett.~{\bf B556}, 169 (2003).
\bibitem{nikitin}
A.~V.~Danilina and N.~V.~Nikitin, 
{\it Four-Leptonic Decays of Charged and Neutral $B$ Mesons within the Standard Model}, 
Phys.~Atom.~Nucl.~{\bf  81}, 347 (2018), Yad.~Fiz. {\bf 81}, 331 (2018);
A.~Danilina N.~Nikitin, and K.~Toms, 
{\it Decays of charged $B$-mesons into three charged leptons and a neutrino}, 
Phys.~Rev.~{\bf D101}, 096007 (2020).
\bibitem{bharucha2021}
A.~Bharucha, B.~Kindra and N.~Mahajan, {\it Probing the structure of the $B$ meson with
$B\to lll'\nu'$}, ArXiv:2102.03193.
\bibitem{beneke2}
M.~Beneke, P.~B\"oer, P.~Rigatos, and K.~K.~Vos, 
{\it QCD factorization of the four-lepton decay $B\to lll\nu$},
Eur.~Phys.~J.~{\bf C81}, 638 (2021). 
\bibitem{wang2022}
C.~Wang, Yu-Ming Wang, Y.-B.~Wei, 
{\it QCD factorization for the four-body leptonic B-meson decays}, 
JHEP {\bf 02}, 141 (2022).
\bibitem{ims}
M.~A.~Ivanov, D.~Melikhov, and S.~Simula, 
{\it Form factors for $B\to j_1 j_2$ decays into two currents in QCD}, 
Phys.~Rev.~{\bf D101}, 094022 (2020). 

\bibitem{bmns2001}
M.~Beyer, D.~Melikhov, N.~Nikitin, and B.~Stech, 
{\it Weak annihilation in the rare radiative $B\to \rho\gamma$ decay}, 
Phys.~Rev.~{\bf D64}, 094006 (2001). 

\bibitem{m2002}
D.~Melikhov, 
{\it Dispersion approach to quark binding effects in weak
decays of heavy mesons},  
Eur.~Phys.~Journal direct {\bf 4}, 2 (2002) [hep-ph/0110087]. 

\bibitem{kruger} 
F.~Kruger and D.~Melikhov,
{\it Gauge invariance and form-factors for the decay $B\to \gamma l^+ l^-$}, 
Phys.~Rev.~{\bf D67}, 034002 (2003).%

\bibitem{kmn2016}
A.~Kozachuk, D.~Melikhov, and N.~Nikitin, 
{\it Annihilation type rare radiative $B_{(s)}\to V\gamma$ decays}, 
Phys.~Rev.~{\bf D93}, 014015 (2016). 

\bibitem{aliev}
T.~M.~Aliev, A.~Ozpineci, and M.~Savci, 
{\it $B_q\to l^+l^-\gamma$ decays in light cone QCD}, 
Phys.~Rev.~{\bf D55}, 7059 (1997).

\bibitem{korchemsky}
G. Korchemsky, D. Pirjol, and T.-M. Yan,
{\it Radiative leptonic decays of $B$ mesons in QCD}, 
Phys.~Rev.~{\bf D61}, 114510 (2000).

\bibitem{kou}
P.~Ball and E.~Kou, 
{\it $B\to \gamma e\nu$ transitions from QCD sum rules on the light cone}, 
JHEP {\bf 0304}, 029 (2003). 

\bibitem{mn2004}
D. Melikhov and N. Nikitin, 
{\it Rare radiative leptonic decays $B_{(d, s)} \to \gamma l^+l^-$}, 
Phys.~Rev.~{\bf D70}, 114028 (2004).

\bibitem{beneke1}
M.~Beneke and J.~Rohrwild,
{\it B meson distribution amplitude from $B\to \gamma l \nu$}, 
Eur.~Phys.~J.~{\bf C71}, 1818 (2011).

\bibitem{kmn2018}
A. Kozachuk, D. Melikhov, and N. Nikitin, 
{\it Rare FCNC radiative leptonic $B_{s,d}\to \gamma l^+l^-$ decays in the Standard Model}, 
Phys.~Rev.~{\bf D97}, 053007 (2018).

\bibitem{beneke2018}
M.~Beneke, V.~M.~Braun, Y.~Ji, and Y.-B.~Wei, 
{\it Radiative leptonic decay $B\to \gamma \ell \nu_\ell$
with subleading power corrections}, 
JHEP {\bf 1807}, 154 (2018). 
\bibitem{ivanov}
S. Dubnicka, A. Z. Dubnickova, M. A. Ivanov, A. Liptaj,
P. Santorelli, and C. T. Tran, 
{\it Study of $B_s\to l^+l^-\gamma$ decays in covariant quark model}, 
Phys.~Rev.~{\bf D99}, 014042 (2019).

\bibitem{zwicky2019}
J.~Albrecht, E.~Stamou, R.~Ziegler, and R.~Zwicky, 
{\it Probing flavoured Axions in the Tail of $B_q\to\mu^+\mu^-$}, 
arXiv:1911.05018.

\bibitem{bobeth}
M.~Beneke, C.~Bobeth, and Y.-M.~Wang, 
{\it $B_{d,s}\to\gamma l^+l^-$ decay with an energetic photon}, 
JHEP {\bf 2012}, 148 (2020).

\bibitem{zwicky2021}
T.~Janowski, B.~Pullin, and R.~Zwicky, 
{\it Charged and neutral $\bar B_{u,d,s}\to \gamma$ form factors
from light cone sum rules at NLO}, arXiv:2106.13616. 





\bibitem{Lattice2021}
A.~Desiderio {\it et al},  
{\it First lattice calculation of radiative leptonic decay rates of pseudoscalar mesons}, 
Phys.~Rev.~{\bf D103}, 014502 (2021).

\bibitem{nachtmann}
D.~Melikhov, O.~Nachtmann, V.~Nikonov, and T. Paulus, 	
{\it Masses and couplings of vector mesons from the pion electromagnetic, 
weak, and $\pi\gamma$ transition form-factors}, 
Eur.~Phys.~J.~{\bf C34}, 345 (2004).

\bibitem{BraunIvanovKorchemsky2004}
V.~M.~Braun, D.~Yu.~Ivanov, G.~P.~Korchemsky,
{\it The B meson distribution amplitude in QCD}, 
Phys.~Rev.~{\bf D69}, 034014 (2004).



\bibitem{ms2000}
D.~Melikhov and B.~Stech,
{\it Weak form-factors for heavy meson decays: an update}, 
Phys.~Rev.~{\bf D62}, 014006 (2000).

\bibitem{ballzwicky2005}
P.~Ball and R.~Zwicky, 
{\it $B_{d,s}\to \rho,\omega, K^*,\phi$ decay form factors from light-cone sum rules reexamined},
Phys.~Rev.~{\bf D71}, 014029 (2005). 

\bibitem{ivanov2} 
M.~A.~Ivanov, J.~G.~K\"orner, S.~G.~Kovalenko, P.~Santorelli, and G.~G.~Saidullaeva, 
{\it Form factors for semileptonic, nonleptonic and rare $B(B_s)$ meson decays}, 
Phys.~Rev.~{\bf D85}, 034004 (2012). 

\bibitem{gubernari2019}
N.~Gubernari, A.~Kokulu and D.~van~Dyk, 
{\it $B\to P$ and $B\to V$ form factors from $B$-meson
light-cone sum rules beyond leading twist}, JHEP {\bf 1901}, 150 (2019). 

\end{thebibliography}
\end{document}